\documentclass[letter]{aa}  

\usepackage{graphicx}
\usepackage{txfonts}
\newcommand\HH{$\mathrm{H_2} $}
\begin{document}

   \title{Analysis of the first infrared spectrum of quasi-bound\\ \HH~line emission in Herbig-Haro 7}


   \author{E. Roueff
          \inst{1}
          \and
                 M. G. Burton\inst{2}
          \and
            T. R. Geballe\inst{3}
          \and
       H. Abgrall\inst{1}
          }

   \institute{Sorbonne Universit\'e, Observatoire de Paris, PSL University, CNRS, LERMA, 
   F-92190,  Meudon, France  \\
              \email{evelyne.roueff@obspm.fr}
         \and
             Armagh Observatory and Planetarium, College Hill, Armagh, BT61 9DB, Northern Ireland\\
             \email {Michael.Burton@Armagh.ac.uk}
            \and
            Gemini Obsevatory/NSF's NOIRLab, 670 N. A'ohoku Place, Hilo, HI 96720, USA\\
             \email{tom.geballe@noirlab.edu}
             }

   \date{Accepted in Astronomy  Astrophysics Letters on december 22, 2022}

 
  \abstract
    {
    Highly excited molecular hydrogen (\HH) has been observed in many regions of shocked molecular gas. A recently published  $K$-band spectrum of Herbig-Haro 7 (HH7) contains several vibration-rotation lines of \HH\ from highly excited energy levels that have not been detected elsewhere, including a line at 2.179 $\mu$m identified as arising from the $v$=2 $J$=29 level, which lies above the dissociation limit of \HH. One emission line at 2.104 $\mu$m in this spectrum was unidentified.}
   {We aim to complete the analysis of the spectrum of HH7 by including previously missing molecular data that have been recently computed.}
   {We re-analysed the $K$-band spectrum, emphasising the physics of  quasi-bound upper levels that can  produce infrared emission lines in the $K$ band.}
   {We  confirm the identification of the $2-1$ $S$(27) line at  2.1785 $\mu$m and identify the line at 2.1042 $\mu$m as due to the 1-0 $S$(29) transition of \HH, whose upper level energy is also higher than the dissociation limit. This latter identification, its column density, and the energy of its upper level further substantiate the existence of a hot thermal component at 5000 K in the HH7 environment.}
   {The presence of the newly identified $1-0$ $S$(29) line, whose quasi-bound upper level ($v$=1, $J$=31) has a significant spontaneous dissociation probability, shows that dissociation of  \HH~is occurring. The mechanism by which virtually all of the \HH~in levels with energies from 20,000 K to 53,000 K is maintained in local thermodynamic equilibrium at a single temperature of $\sim$5,000 K remains to be understood.}


   \maketitle
%

\section{Introduction}
\label{sec:introduction}
   The interaction of the collimated outflow from the protostar SSV13 \citep{strom:76} and the molecular cloud out of which it formed has produced a collection of Herbig-Haro (HH) objects, HH7-HH11, in a more or less linear arrangement on the sky. The most distant of these from SSV13, HH7, has a  classic bow shock shape.  It is bright in line emission from  shock-excited vibrational states of molecular hydrogen (\HH), first observed  in  
the  $v=1-0$ $S$(1) transition by \cite{zealey:84},  {\cite{hartigan:89}, and \cite{garden:90} and subsequently in vibrational levels $0-4$ and rotational levels $1-15$ by   \cite{burton:89}  and  \cite{fernandes:95}. HH7 also emits strongly in  pure rotational lines of \HH~and CO  \citep{neufeld:06a, yuan:11,neufeld:19,molinari:00} as well as in [OI]$_{63\mu}$ \citep{sperling:20}, H$_{\alpha}$, [OI]$_{\lambda6300}$, and [SII]$_{\lambda6716}$ \citep{hartigan:19}. The vibrationally excited \HH~lines, observed mainly in the $2.0-2.5$ $\mu$m interval, are emitted predominantly in the hottest shock-heated gas, while the pure rotational low-$J$ transitions of \HH~and the pure rotational transitions of CO, observed in the mid- and far-infrared, arise  in a somewhat cooler gas downstream. 

Much more highly vibrationally and rotationally excited molecular hydrogen was found  by \citet[][hereafter P16]{Pike:16} in a  3\arcsec $\times$ 3\arcsec region near the tip of the HH7 bow shock, in
  $K$-band spectra they obtained at a resolving power, R, of 5000. Figure 6 of their
   paper shows the $2.01-2.45~\mu$m spectrum of a 0\farcs6$\times$0\farcs9 area in that region. Their paper demonstrated the existence in HH7 of a small 
  percentage (1.5\%) of the emitting \HH\ at a temperature of $\sim$5,000 K.  Subsequently, \cite{geballe:17} discovered the presence of  small percentages  of 5,000 K 
  \HH\ in shocked gas at several locations in the Orion Molecular Cloud. \cite{giannini:15} detected a similar phenomenon in another bright HH object, HH1, at a somewhat higher temperature, $\sim$6300 K.
 
   P16 identified a weak emission line at 2.179 $\mu$m in the HH7 spectrum as the $2-1$ $S$(27) transition of \HH, which 
  arises from the upper level, $v=2, J=29$, whose energy is above the dissociation limit of the ground state of \HH; this corresponds to
  51,965.84 K, using the latest measurements \citep{Holsch:19} and the Committee on Data for Science and Technology (CODATA) definition of fundamental constants \citep{Tiesinga:21}.
 The column density of \HH\ in the upper level could only be crudely estimated by  P16, as the Einstein A coefficient for that transition was not known.
  P16  also  reported the detection of a faint line near 2.104 $\mu$m, which they were unable to identify. 

   \cite{Roueff:22} have recently proposed a simple and efficient method for computing  the emission spectrum produced by quasi-bound levels,
providing accurate wavenumbers and Einstein emission coefficients.  The application to \HH ~allowed them to calculate the Einstein coefficient of 
the 2-1 $S$(27) transition and suggested that the line at 2.104  $\mu$m in HH7 is the $1-0$ $S$(29) transition of \HH, whose upper-state energy 
also lies above the dissociation limit.

The present paper analyses these two high excitation lines in the light of the  new theoretical developments. Section \ref{sec:obs} revisits the observations of HH7, Sect. \ref{sec:theory} summarises the recent theoretical achievements, and Sect. \ref{sec:analysis} contains the resulting extended observational analysis of the \HH\ line emission in HH7. We provide a discussion of our results in Sect. \ref{sec:discussion}. 
 \begin{figure}
 \includegraphics[width=8.8cm]{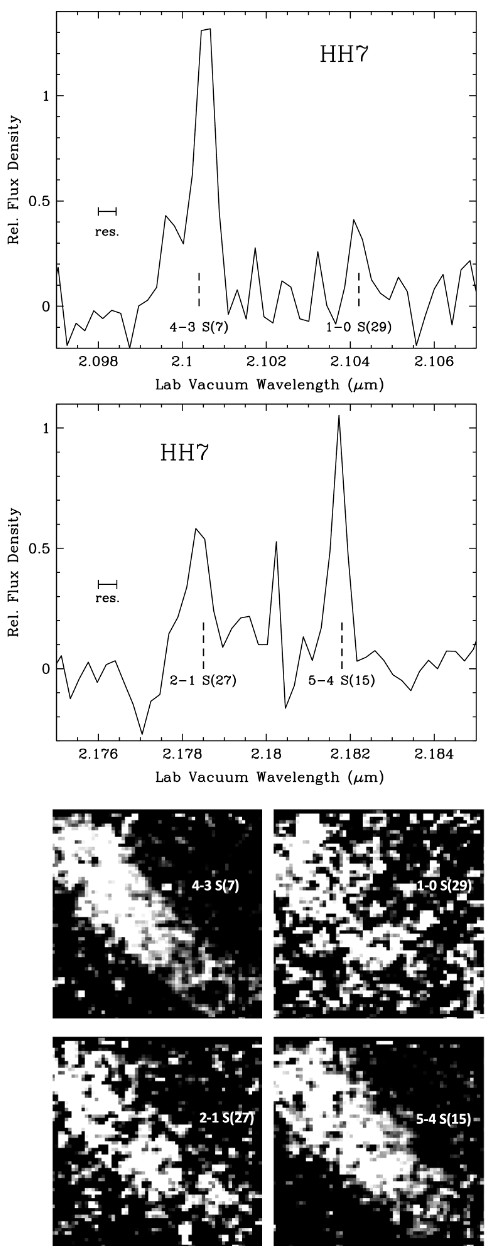}
 \caption{Observational data showing highly excited \HH\ lines in HH7.
 Top two panels: Spectra of a 0\farcs6 $\times$ 0\farcs9 area of HH7 in two narrow wavelength intervals, each containing a line of \HH\ from a quasi-bound energy level and one adjacent line, from Fig. 6 of P16. Vertical dashed lines are the line wavelengths calculated as described in Sect. \ref{sec:theory}. Bottom: Spectral images of the four lines shown above, extracted from the NIFS data cube. The field of view is 2\farcs5 $\times$ 2\farcs5 and corresponds to the left part of Fig. 2 of P16; the field centre corresponds to RA = 3:29:08.42, Dec = +31:15:27:45 (J2000), with an estimated uncertainty of 0$\farcs$25. } 
 \label{fig:spectra}
 \end{figure}
 
 \section{Observations}
 \label{sec:obs}
 
 A detailed description of the observations of HH7 and a reduction of the spectral data have been given by  P16. In brief, the Gemini facility integral field spectrometer, 
 the  Near Infrared Field Spectrometer (NIFS; \citealt{McGregor:03}), was used at the \textit{Frederick C. Gillett }Gemini North Telescope on Maunakea, Hawai'i, to obtain
 spectra of a 3\arcsec $\times$ 3\arcsec region near the tip of the HH7 bow shock, for program GN-2007B-Q-47.  The angular resolution of the spectra was 0\farcs35.  Within this 3\arcsec $\times$ 3\arcsec region, the spectra showed \HH\ ro-vibrational line emission from upper-state levels covering a wide range of energies, including a dozen in the range $40,000-50,000$ K.  Because some rotational  energies and associated rotational quantum numbers of the upper levels of these lines are high ($J$ $\gtrsim$ 15), collisions rather than the absorption of ultraviolet (UV) photons are probably the main producer of the populations in those rotational levels. Somewhat lower values of $J$ associated with high vibrational quantum numbers are commonly found in  dense photon-dominated regions (PDRs) such as NGC 2023, the Orion Bar, S140, and IC63.  \citep{burton:92,mccartney:99, kaplan:21}. \HH\  is   excited in PDRs  by UV pumping, which is followed by electronic fluorescence, but  the $\Delta J = \pm 1$ selection rule for electronic transitions maintains $J$ at values below $\sim$ 13.\footnote{We  contacted K. Kaplan to check if the two transitions at 2.1785 $\mu$m and 2.1042 $\mu$m were present in his PDR spectra obtained with the Immersion Grating INfrared Spectrometer (IGRINS), and they were not.}  
 
P16 concentrated their analysis on the spectrum of the 0\farcs6 $\times$ 0\farcs9 area shown in their Fig. 2; the spectrum is plotted in their Fig. 6.  The upper two panels of our Fig. \ref{fig:spectra} show in more detail two 0.01~$\mu$m wide portions of that spectrum, each containing one of the two highly shock-excited \HH\ lines discussed in the Introduction. Wavelength calibration employed the spectrum of an argon lamp and is accurate to $\sim$ 0.00002~$\mu$m. The horizontal scales are vacuum laboratory wavelengths and as such can be directly compared with the theoretically calculated wavelengths (see Sect. \ref{sec:theory}).
The uppermost  panel contains the previously unidentified line at 2.1042~$\mu$m along with the nearby $4-3$ $S$(7) line. Similarly, the middle panel contains the previously identified $2-1$ $S$(27) line and the adjacent $5-4$ $S$(15) line.  Spectral images of the four lines, extracted from the NIFS data cube, are shown in the bottom panel of the figure and demonstrate that, to within the limits imposed by the noise levels, the four emission lines have identical morphologies, which also match the morphology of the strong $1-0$ $S$(1) line shown in Fig. 2 of P16. Based on the fluctuations in the baseline, we estimate the confidence of the detection of the $1-0$ $S$(29) line to be 3.5$\sigma$. The wavelengths of these two weak lines are slightly different than those reported by P16 and are more accurate. 
\section{Theoretical aspects}
\label{sec:theory}

Molecular quasi-bound levels correspond to  states whose energies lie above the dissociation limit of the ground state of the molecule but   well  below the dissociation energy of the  electronically  excited molecule. For \HH, the Schr{\"o}dinger equation relevant to excited rotational levels
is
 \begin{equation}
 \label{eq:schr}
- \frac{\hbar^2}{2 \mu}  \cdot  \frac{d^2f_{v,J}(R)}{dR^2}  + V_{eff}^{mod}(R,J) f_{v,J}(R) =  E_{v,J} f_{v,J}(R),
\end{equation} 
where 
$V_{eff}(R,J) = V(R) + \frac{\hbar^2 J\,(J+1)}{2 \mu R^2}$, 
with $V(R)$ the ground state electronic molecular potential of \HH ~and $\mu=M_\mathrm{p}/2$ the nuclear reduced mass of \HH. 
 
   \begin{figure}[h]
\includegraphics[width=8cm]{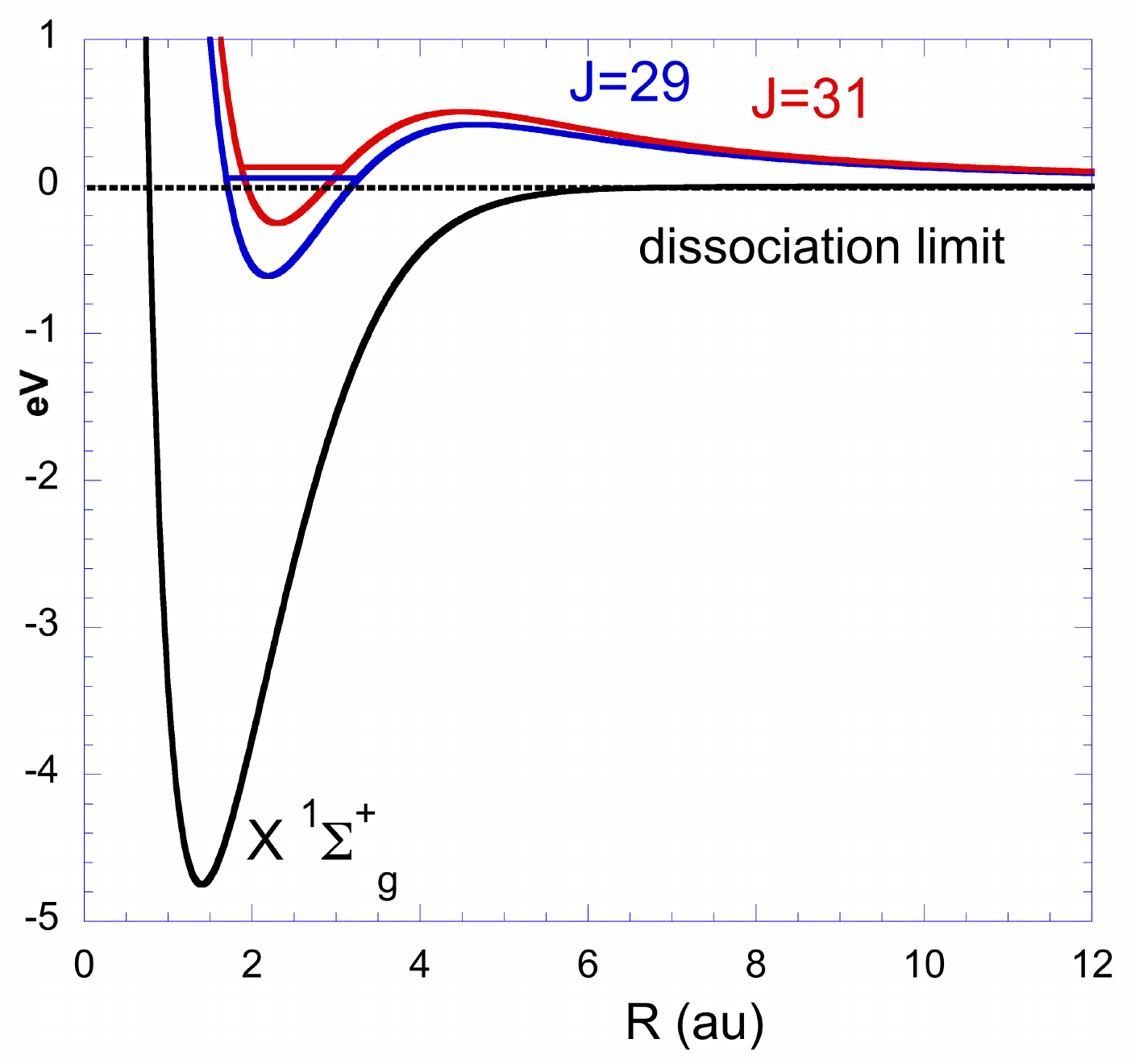}
   \caption{\HH ~molecular potentials in eV as a function of the interatomic distance, $R$, expressed in atomic units. The zero value  
   corresponds  to photo-dissociated \HH.  The black curve is the electronic potential of the X $^1\Sigma_g^+$ ground state
   from \cite{Komasa:18} expressed in eV.
   The blue and red curves denote the effective potentials with $J$=\,29 and $J$=\,31, respectively. 
   The quasi-bound levels $v=2, J=29$ and $v=1, J=31$ are also  
   displayed in blue and red, respectively, in the allowed ranges of interatomic distances.}
              \label{fig:h2}%
    \end{figure}

Figure \ref{fig:h2} displays the electronic molecular potential of the X$^1\Sigma_g^+$ ground state of \HH ~as well as the effective potentials corresponding to $J=29$ and $J=31$, the two quasi-bound levels (sometimes referred to as shape resonances) previously mentioned. The presence of the centrifugal potential,  $ \frac{\hbar^2 J\,(J+1)}{2 \mu R^2}$,  significantly modifies the shape of the electronic contribution, $V(R)$, by reducing the potential well, shifting the minima to larger interatomic distances and exhibiting broad bump maxima above the dissociation limit, peaking near 4.5 atomic units.

Figure \ref{fig:h2} also displays the resonant quasi-bound eigenvalues,  $E_r$, which are located above the dissociation limit and are trapped inside the centrifugal barrier. The associated wave function for each level has a non-vanishing probability in the interatomic range displayed, becomes vanishingly small after the second turning point when $E_r  \le V_{eff}(R)$, and has an oscillatory behaviour for large R when  $E_r$  becomes larger than $V_{eff}(R)$.
The associated quasi-discrete stationary states have complex energy eigenvalues, $E= E_r-(i \, \Gamma/2),$ where $E_r$ is the energy at resonance and $\Gamma$ characterises the width of the level and determines its lifetime against dissociation, $\tau =  \hbar  / \Gamma$, due to tunnelling from the quasi-bound to 
the continuum oscillatory dissociating state at large interatomic distances. \cite{Roueff:22} computed the various resonance energy level positions of \HH\ and  the corresponding emission spectrum arising from these levels by using the recent highly accurate molecular potential of the \HH~ground state of \cite{Komasa:18} and extending the effective potential by a constant value from the maximum value of the potential function. This method allows one to use a standard numerical integration of the Schr{\"o}dinger equation applied to strictly bound levels and has been demonstrated to be very precise for determining the resonant energy level positions and the emission rates. However, it does not allow  a derivation of the widths or the dissociation lifetimes. Those are obtained through different methods based on scattering properties \citep{Schwenke:88,Selg:10}. 

These computations predict wavelengths of 2.1785  $\mu$m for the $2-1$ $S$(27) transition and 2.1042  $\mu$m for the $1-0$ $S$(29) transition.  The predicted wavelengths for the two stronger lines in Fig. \ref{fig:spectra} are 2.10043 $\mu$m for $4-3$ $S$(7) and 2.18179 $\mu$m for  $5-4$ $S$(15). As can be seen in the figure, all are in excellent agreement with the observed wavelengths. Therefore, we are confident in the previous identification of the $2-1$ $S$(27) line by  P16  and in our 
identification of the weak and previously unidentified feature at 2.1042 $\mu$m as the $1-0$ $S$(29) line.   These two transitions are the only lines in the 2.01 - 2.45 $\mu$m interval from quasi-bound levels that would have been detectable in our data. (We note in Table 1 the small Einstein A coefficient of the $2-0$ $Q$(29) line at 2.4007 $\mu$m.)

 \section{Column density analysis}
 \label{sec:analysis}
The analysis undertaken here follows that described in P16  for the H$_2$ line emission from HH7 reported in that paper, with the addition of the 1--0 $S$(29) and 2--1 $S$(27) lines presented here.
A two-component Boltzmann distribution with temperatures $T_{hot}$ and $T_{warm}$ was fitted to the column densities obtained from the de-reddened line intensities,
\begin{equation}
N_i = N_{i,hot} + N_{i,warm}
\label{eqn:fitformula}
,\end{equation}
with each component described by a Boltzmann distribution at the corresponding temperatures, as per  P16.  This is shown in Fig.~\ref{fig:coldenfit}.

We obtain  $T_{warm} = 1,783 \pm 20$\,K and $T_{hot} = 5,133 \pm 17$\,K, with 98.5\% of the total column of excited H$_2$ gas in the warm component of the gas and 1.5\%  in the hot component.  This compares to values of $T_{warm} = 1,803 \pm 12$\,K and $T_{hot} = 5,200 \pm 12$\,K found without these two extra lines included in the analysis\footnote{The errors quoted here are the formal errors derived from the least squares fit.}.   The additional lever arm provided by the two higher excitation energy levels has only led to a marginal decrease in the derived temperatures; in other words, the result is essentially the same.

We conclude that the two quasi-bound H$_2$ lines are well modelled by the same hot local thermodynamic equilibrium (LTE) component as per all lines measured in HH7 arising from energy levels $\geq$15,000\,K\@.  The level populations for the two quasi-bound lines are $\sim10^{-5}$ times that of the $v=1$, $J=3$  upper level of the brightest H$_2$ emission line, $1-0$ $S$(1).


\begin{figure}
\includegraphics[width=9.0cm]{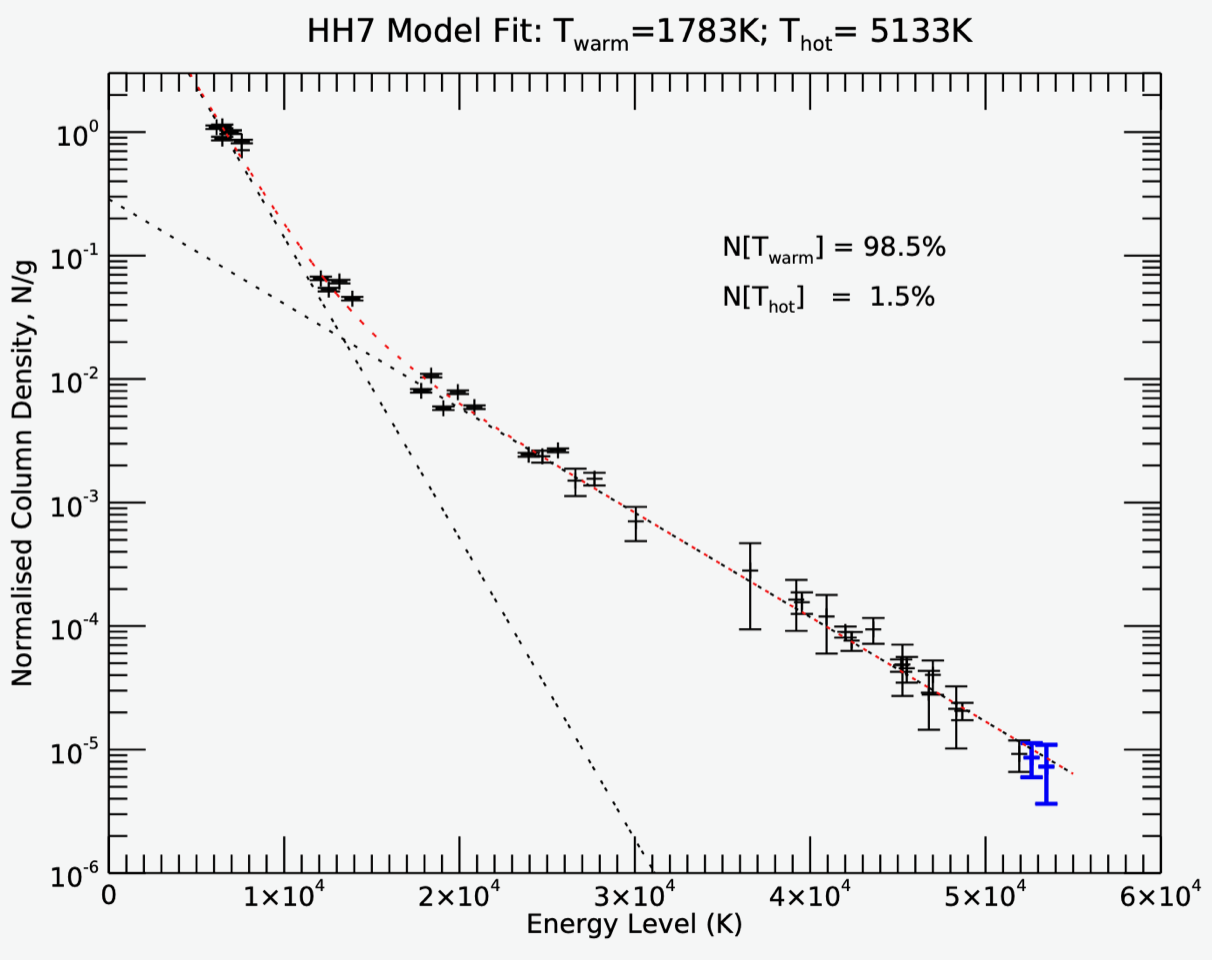}
\caption{Level column densities, divided by their degeneracies, $N_i / g_i$, plotted as a function of level energy, $T_i$, for the H$_2$ lines measured in HH7\@.  They are normalised to unity for the $(v,J)=(1,3)$ upper-state level at 6,952\,K, which emits the 1--0 $S$(1) line.  The two blue points (in the lower right) are for the newly analysed  2--1  $S$(27) and  1--0  $S$(29) lines.   The dashed red line shows the best two-temperature LTE fit, as described in Sect. \ref{sec:analysis}.  
\label{fig:coldenfit}}
\end{figure}

\begin{table*}[h]
\caption{Properties of the two detected quasi-bound levels, $v=2, J=29$ and $v=1, J=31$, of H$_2$ and their emission spectrum.}
\centering
\begin{tabular}{lccccccc}  
\hline\hline
Transition& $\tilde{\nu}$ & $\lambda$ &   $A$  &  $A_r$    &  $\tau_d$ &$E_{upper}^{qb}$ \\ 
 label          &   cm$^{-1}$     &      $\mu$m         &         s$^{-1}$ &   s$^{-1}$   &   s    & K  \\ 
 \hline
    $2-0$ $O$(31) &  1387.04  &  7.2096  &     2.704E-11     & 5.482E-06  & 8.130E12  & 676.0  \\ 
   $2-0$ $Q$(29) &   4165.40 &   2.4007 &  4.592E-08    &  5.482E-06   &   8.130E12   &   676.0\\ 
  $2-0$ $S$(27) &  7034.45 &   1.4216  &   2.240E-07     &   5.482E-06  &  8.130E12  &  676.0    \\ 
  $2-1$ $Q$(29) &  1944.67  &    5.1423&     3.947E-07    &  5.482E-06   &   8.130E12 &   676.0  \\ 
 $2-1$ $S$(27) &   4590.29  &   2.1785 &  2.944E-06    &  5.482E-06  &   8.130E12  &    676.0    \\ 
$2-2$ $S$(27)&    2399.19  &    4.1681 &   1.873E-06     &   5.482E-06    &   8.130E12  & 676.0   \\ 
 $2-3$ $S$(27) &  489.60 &  20.4248 &    2.582E-10      &   5.482E-06   &  8.130E12  &  676.0  \\ 
   $1-0$ $Q$(31) &  1974.02  & 5.0658  &  2.452E-07     & 5.219E-06  & 4.083E06   &   1520.5   \\ 
   $1-0$ $S$(29)&    4752.38   & 2.1042   &   2.101E-06        &   5.219E-06&  4.083E06    & 1520.5    \\ 
  $1-1$ $S$(29) &  2531.65     & 3.9500  &     2.872E-06       &   5.219E-06& 4.083E06    &  1520.5   \\ 
  $1-2$ $S$(29) &    586.98  & 17.0364  &   4.963E-10     &   5.219E-06 & 4.083E06   &  1520.5    \\ 
 \hline
  \end{tabular}
\tablefoot{$\tilde{\nu}$  is the computed transition   frequency;  $\lambda$ is the corresponding  vacuum wavelength. A is the sum of the electric quadrupole   and 
 the magnetic dipole contributions to the Einstein radiative emission coefficients of the transition from \cite{Roueff:22}.  $A_r$ is the total radiative decay probability, and $\tau_d$ is the dissociation lifetime of the upper level. $E_{upper}^{qb}$ is the
quasi-bound upper level energy expressed in K above the dissociation limit.} 
\label{tab:res}
\end{table*}

\section{Discussion}
\label{sec:discussion}

Table \ref{tab:res} summarises the present knowledge available for the two quasi-bound levels of \HH\, $v$ = 2, $J$ = 29 and 
$v$ = 1, $J$ = 31, that have been detected. 
The upper level involved in the $2-1$ $S$(27) transition at 2.1785 $\mu$m,  676 K above the dissociation energy of the ground state, is very stable  against dissociation,  whereas  that of the 
 $1-0$ $S$(29) transition at 2.1042 $\mu$m,  located 845 K higher, has a dissociation probability of   approximately five percent 
 and a dissociative lifetime, $ \tau_d  = \hbar /  \Gamma_d$, resulting from quantum tunnelling through the centrifugal barrier (see Fig. \ref{fig:h2}) of  4.083 $\times~ 10^6$ s, corresponding to less than two months.  This indicates that the shock wave in HH7 is partially dissociative. 


As shown in Fig.~\ref{fig:coldenfit}, the  5,000 K component represents a small percentage of the line-emitting \HH\ in HH7. As noted previously, similar small percentages have been observed in HH1 and in the Orion molecular outflow.  Figure~\ref{fig:coldenfit} also shows that \HH\ in energy levels greater than $\sim$ 20,000 K above   the ground state are populated only by this component.  In the case of HH1, \cite{giannini:15} observed a wide range of neutral and ionised species emitting in close proximity to the \HH, many at optical wavelengths. Their analysis yields a temperature range of $8,000-80,000$ K to account for the emission.  They further find that neutral and fully ionised regions coexist inside the shock. However, for the heavily extincted \HH~line emission from HH7 ($A_{\textrm{V}}$ = $12-28$ mag;  P16), the species producing the optical emission lines observed by \citet{solf:87},  \cite{hartigan:89}, and \cite{ hartigan:19} cannot be mixed with the \HH.  


In view of the detections by \cite{giannini:15},  P16, and \cite{geballe:17} of  $5,000-6,000$ K \HH~in diverse environments, it seems likely 
that  a small percentage of \HH~existing at those temperatures is a common occurrence in collisionally shocked molecular gas, at least in cases where collisions between 
outflows and ambient molecular material occur at velocities of many tens of km s$^{-1}$, as is the case for HH1, HH7, and the Orion Molecular Cloud. 
  In addition, although transitions emitted from quasi-bound levels have only been detected towards HH7, we expect that they are present in HH1 and OMC-1 at roughly the same intensities relative to the stronger \HH\ lines, as in HH7. 

 
 It is generally accepted that the maximum temperatures  of nearly all of the vibrationally excited \HH~in each of the above shocked clouds and in many others are suppressed by continuous shocks, in which the collisional acceleration of the ambient clouds and deceleration of the colliding outflows from the protostars are sufficiently gradual to heat the \HH\ only to temperatures of $\sim$2,000 K and prevent its dissociation (for more details, see Sect. I of  P16 and references therein).  The existence of \HH\ at a range of lower temperatures in gas cooling behind the continuous shocks, which has been demonstrated by observations of pure rotational lines \citep[e.g.][]{neufeld:19}, is also unsurprising.  However, it seems remarkable that virtually all of the highly ro-vibrationally excited \HH\ in levels with energies from 20,000 K to 53,000 K is maintained in LTE at a single temperature of $\sim$5,000 K, and that there is virtually no \HH\  at temperatures between 2,000 K and 5,000 K.  The mechanism that produces this bimodal temperature distribution is unclear.   

The location and morphology of the 5,000 K gas also is unclear. The gas could be located in thin (currently unresolvable) sheets where the molecular cloud is being collisionally accelerated, the wind is being collisionally decelerated, or both. Its line emission could alternatively also be occurring in small clumps of unusually hot and/or unusually dense gas scattered along the shock front. Comparisons of the velocity profiles of lines originating from levels whose populations are dominated by the gas at 5,000 K with those from levels dominated by the 2,000 K component, at higher spectral resolution than has been employed to date, might reveal small differences and constrain the relative locations of the two components. The good fit of the $v$=1, $J=31$ column density to the fit to the population-energy diagram (Fig.~\ref{fig:coldenfit}) indicates that  dissociation is taking place in the 5,000 K gas.

 One can consider if the short lifetime of the v=1, J=31 quasi-bound level indicates a significant continuous reformation of molecular hydrogen in the gas phase at high temperatures. We have estimated the formation rate of H$_2$ through radiative association via that resonance level, $i$,  $ {\rm{H + H  \leftrightarrow  H_2}}^i \rightarrow H_2 + h \nu$, ~following the theory of \cite{Bain:72}, to be
\begin{equation}
\alpha_i^{res}= \left( \frac{2 \pi \hbar^2}{M k T} \right)^{3/2} {(2 I + 1)} (2J_i+1) \,\,\frac{A_{r}^i \ A_d}{A_{r} ^i+A_d} \,\,exp(-E_i/kT),
\end{equation}
 where $A_d = 1/ \tau$ and $M$ is the reduced mass  of the colliding atoms. 
The contribution of $v=1$, $J=31$ with I = 1 and similar values of $A_r$ and $A_d$ is the most efficient by orders of magnitude.
However, the derived value for its contribution at 5000K is 1.37 $\times$ 10$^{-30}$ cm$^3$ s$^{-1}$, which is negligible. 

 Although it is difficult to assess the direct implication of the measurable presence of these quasi-bound \HH\ states for shock chemistry, their detections confirm the predictability of theoretical computations based on highly accurate potential curves. The physical conditions associated with the astrophysical environments in which their lines are emitted may not be reproducible in the laboratory due to their very large rotational quantum numbers. Thus, they probably offer the only way to probe these levels. 
\cite{martin:96} introduced quasi-bound levels of \HH\ in their master equation studies of collisional excitation of \HH\ by H
and specifically mentioned the $v=2$, $J=29$, and $v=1$, $J=31$ quasi-bound levels detected here. However, they find that the highly excited rotational levels are not thermally populated for the range of physical conditions that they considered, in contrast to what astronomical observations have revealed.  Finally, we note that the contribution of quasi-bound levels to the partition function of \HH~and its isotopologues has been recently computed by \cite{zuniga:21} using the same potential as us. 




\begin{acknowledgements}
      We thank K. Kaplan for having searched the two transitions in his IGRINS spectra of various PDRs. We are grateful to the referee for helpful comments.  E.R. and H.A. acknowledge support by the Programme National de Physique et de Chimie du Milieu Interstellaire (PCMI) of  CNRS/INSU with INC/INP co-funded by CEA and CNES.This research is based in large part on observations obtained at the international Gemini Observatory, a program of NSF's NOIRLab, which is managed by the Association of Universities for Research in Astronomy (AURA) under a cooperative agreement with the National Science Foundation, on behalf of the Gemini Observatory partnership: the National Science Foundation (United States), National Research Council (Canada), Agencia Nacional de Investigaci\'{o}n y Desarrollo (Chile), Ministerio de Ciencia, Tecnolog\'{i}a e Innovaci\'{o}n (Argentina), Minist\'{e}rio da Ci\^{e}ncia, Tecnologia, Inova\c{c}\~{o}es e Comunica\c{c}\~{o}es (Brazil), and Korea Astronomy and Space Science Institute (Republic of Korea).  
  \end{acknowledgements}


\end{document}